\documentclass[11pt,a4paper,final,standalone]{iopart}

\usepackage{tikz}
\usetikzlibrary{mindmap,trees}
\usepackage{xcolor}
\definecolor{myOrange}{RGB}{0, 104, 150}
\definecolor{myRed}{RGB}{196, 49, 25}
\definecolor{myGreen}{RGB}{63, 126, 49}
\definecolor{myPlum}{RGB}{135, 41, 150}
\usepackage[utf8]{inputenc}
\usepackage[T1]{fontenc}
\usepackage{color} 
\usepackage{graphicx}
\usepackage{bm}
\usepackage{rotating}
\usepackage{multirow}

\usepackage{hyperref}
\usepackage[normalem]{ulem}
\usepackage{iopams}
\usepackage{titlesec}

\expandafter\let\csname equation*\endcsname\relax
\expandafter\let\csname endequation*\endcsname\relax

\usepackage{amsmath} 
\usepackage{amssymb} 

\def\orcid#1{\kern .08em\href{https://orcid.org/#1}{\includegraphics[keepaspectratio,width=0.7em]{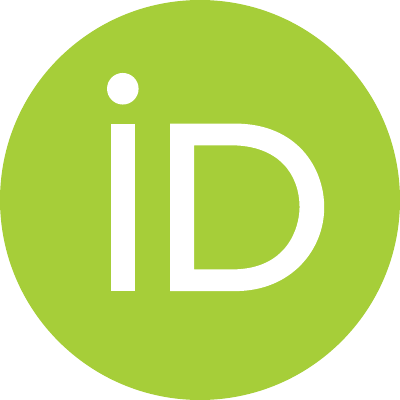}}}

\usepackage{caption}
\usepackage{pifont}

\begin{document}

\newcommand{\IR} [1]{\textcolor{red}{#1}}

\title{Paths to Superheavy Nuclei}

\author{K. Godbey$^{1,*}$, F.M. Nunes$^{1,2}$,
M. Albertsson$^{3,4}$,
K.J. Cook$^{1,5}$,
J.M. Gates$^{3}$,
K. Hagel$^{6}$,
K. Hagino$^{7}$,
M. Kowal$^{8}$,
Jin Lei$^{9}$,
J. Lubian$^{10}$,
A. Makowski$^{11}$,
P. McGlynn$^{1}$,
M. R. Mumpower$^{12}$,
W. Nazarewicz$^{1,2}$,
G. Potel$^{13}$,
J.L. Pore$^{3}$,
J. Rangel$^{14}$,
K. Sekizawa$^{15,16,17}$,
A.S. Umar$^{18}$
}

\address{$^{*}$E-mail: godbey@frib.msu.edu}
\address{$^1$ Facility for Rare Isotope Beams, East Lansing, MI 48824, USA.}
\address{$^2$ Department of Physics and Astronomy, Michigan State University, East Lansing, MI 48824-1321, USA.}
\address{$^{3}$ Nuclear Science Division, Lawrence Berkeley National Laboratory, Berkeley, California 94720, USA}
\address{$^{4}$ Division of Mathematical Physics, Lund University, 221\,00 Lund, Sweden}
\address{$^{5}$ Department of Nuclear Physics and Accelerator Applications, The Australian National University, Canberra, ACT 2601, Australia}
\address{$^{6}$Cyclotron Institute, Texas A \& M University, College Station, Texas 77843 USA}
\address{$^{7}$ Department of Physics, Kyoto University, Kyoto 606-8502, Japan }
\address{$^{8}$ National Centre for Nuclear Research, Pasteura 7, 02-093 Warsaw, Poland }
\address{$^{9}$ School of Physics Science and Engineering, Tongji University, Shanghai 200092, China.}
\address{$^{10}$ Istituto de Física, Universidade Federal Fluminense, Av. Litorânea, Gragoatá, Niterói, R.J. Brazil 24210-340}
\address{$^{11}$ Faculty of Physics, Warsaw University of Technology, Ulica Koszykowa 75, 00-662 Warsaw, Poland}
\address{$^{12}$ Theoretical Division, Los Alamos National Laboratory, Los Alamos, NM 87545}
\address{$^{13}$ Departamento de F\'isica Aplicada III, Escuela T\'ecnica Superior de Ingenier\'ia, Universidad de Sevilla, Camino de los Descubrimientos, Sevilla, Spain}
\address{$^{14}$ Departamento de Matem\'atica, F\'{i}sica e Computa\c{c}\~{a}o Universidade do Estado do Rio de Janeiro, Faculdade de Tecnologia, 27537-000, Resende, Rio de Janeiro, Brazil}
\address{$^{15}$ Department of Physics, School of Science, Institute of Science Tokyo, Tokyo 152-8551, Japan }
\address{$^{16}$ Nuclear Physics Division, Center for Computational Sciences, University of Tsukuba, Ibaraki 305-8577, Japan}
\address{$^{17}$ RIKEN Nishina Center, Saitama 351-0198, Japan}
\address{$^{18}$ Department of Physics and Astronomy, Vanderbilt University, Nashville, TN, 37235, USA}

\date{\today}

\newcommand{\cmark}{\ding{51}}%
\newcommand{\xmark}{\ding{55}}%
\newcommand {\nc} {\newcommand}


\newcommand{\ket}[1]{| #1 \rangle}
\newcommand{\bra}[1]{\langle #1 |}

\newcommand{\ie}{\emph{i.e.}}
\newcommand{\eg}{\emph{e.g.}}
\newcommand{\etc}{\emph{etc.}}

\newcommand{\nuc}[2]{$^{#1}$#2}

\maketitle
\begin{abstract}
This document summarizes the discussions and outcomes of the Facility for Rare Isotope Beams Theory Alliance (FRIB-TA) topical program “The path to Superheavy Isotopes” held in June 2024 at FRIB. Its content is non-exhaustive, reflecting topics chosen and discussed by the participants. The program aimed to assess the current status of theory in superheavy nuclei (SHN) research and identify necessary theoretical developments to guide experimental programs and determine fruitful production mechanisms. This report details the intersection of SHN research with other fields, provides an overview of production mechanisms and theoretical models, discusses future needs in theory and experiment, explores other potential avenues for SHN synthesis, and highlights the importance of building a strong theory community in this area.

\end{abstract}

\tableofcontents

\newpage
\markboth{}{}

\section{Introduction}
\label{sec:intro}

Superheavy nuclei (SHN), or nuclides having proton number $Z \geq 104$, represent extremes of nuclear matter. Complementary to studies of nuclear structure near the proton and neutron driplines, the synthesis and study of new superheavy nuclides provide stringent tests of models of nuclear structure and reactions. Research into superheavy nuclides impacts our understanding of quantum many-body systems, nuclear astrophysics, and tests the extremes of atomic physics and chemistry. Many open questions remain in this domain \cite{lrp2023,smits2024}, including foundational questions such as: does the periodicity of chemical properties end?

A critical prediction regarding the properties of SHN is the existence of a region of enhanced stability near $Z\sim~114$, $N\sim~184$. Here, half-lives are expected to significantly increase, leading to a region of enhanced stability \cite{Nazarewicz2018,smits2024}. However, the exact degree of stability of these SHN is unknown. While we have likely already reached the neutron-deficient `shore' of this peninsula of stability, experimental advances guided by theory will be required to understand its full extent and the properties of the isotopes within it.

The primary mechanism for producing superheavy elements in the laboratory has been the fusion of stable beams with long-lived targets. Elements 113-118 were synthesized via hot fusion reactions of \nuc{48}{Ca} impinging on increasingly heavier actinide targets. This hot-fusion method faces obvious limitations. The heaviest element produced, Oganesson ($Z=118$), was synthesized with a radioactive target: \nuc{48}{Ca} + \nuc{249}{Cf} ($Z=98$). The short half-life of the next heavier actinide, Es ($Z=99$), precludes its use as a target for SHN production. Projectiles heavier than \nuc{48}{Ca}, such as $^{50}$Ti~\cite{gates2024,Oganessian2025}, $^{51}$V~\cite{sakai2022facility}, and $^{54}$Cr~\cite{Oganessian2025}, are also being considered, although evaporation residue yields rapidly decrease with increasing projectile charge. Another production method considered is the use of neutron-rich rare isotopes, but here too the gain in cross section is not sufficient to compensate for the much reduced intensity of the beams. Similar to generating new superheavy elements (increasing $Z$), synthesizing increasingly neutron-rich isotopes of existing superheavy elements (increasing $N$) has proved extremely challenging because stable-beam fusion with long-lived targets produces neutron-deficient SHN. New reaction pathways are clearly needed.

While experimental searches for elements 119 and 120 are underway, significant progress requires intensive theoretical and experimental efforts. Several open questions remain in fusion, including those related to the ‘optimality’ of a given entrance channel. At the same time, exploring novel pathways to element formation, such as multinucleon transfer and leveraging insights from reaction theory, is important to diversify efforts and maximize discovery potential. This paper addresses the current limitations in theory for predicting SHN properties and explores promising efforts in both theory and experiment.

This document summarizes the discussions and outcomes of the Facility for Rare Isotope Beams Theory Alliance (FRIB-TA) topical program “The path to Superheavy Isotopes” held in June 2024 at FRIB. Its content is non-exhaustive, reflecting topics chosen and discussed by the participants. The goal of the topical program was to assess the current status of theory and discuss which theory developments are needed to guide the experimental program and determine the most fruitful production mechanisms. We begin in Sec.~\ref{sec:motivation} by discussing the many intersections with other fields. In Sec.~\ref{sec:production}, we provide a broad overview of how SHN production is understood. Next, we summarize many of the current models used in this field (Sec.~\ref{sec:models}), followed by a discussion of future developments needed (Sec.~\ref{sec:needs}). The experimental context is provided in Sec.~\ref{sec:experiments}, and Sec.~\ref{sec:other} includes other avenues discussed in the topical program. In closing, we discuss the need to strengthen the theory community in this area (Sec.~\ref{sec:community}).

\section{Motivation and intersections with other fields}
\label{sec:motivation}

Research in superheavy nuclei and atoms connects many areas of science. Numerous examples of such interdisciplinary intersections can be found in recent reviews \cite{Nazarewicz2018,Giuliani2019,Smits2023,smits2024}. Some are discussed below.

\subsection{Physics of complex systems}
Due to their large number of nucleons, superheavy nuclei are approaching the leptodermous limit \cite{Reinhard2006} in which shell effects are suppressed \cite{Bender2001,Jerabek2018} and the structure is primarily driven by a bulk nuclear matter description. In addition, large electric fields give rise to Coulomb frustration effects.

\begin{figure}[tbh!]
\centering
\includegraphics[width=0.8\linewidth]{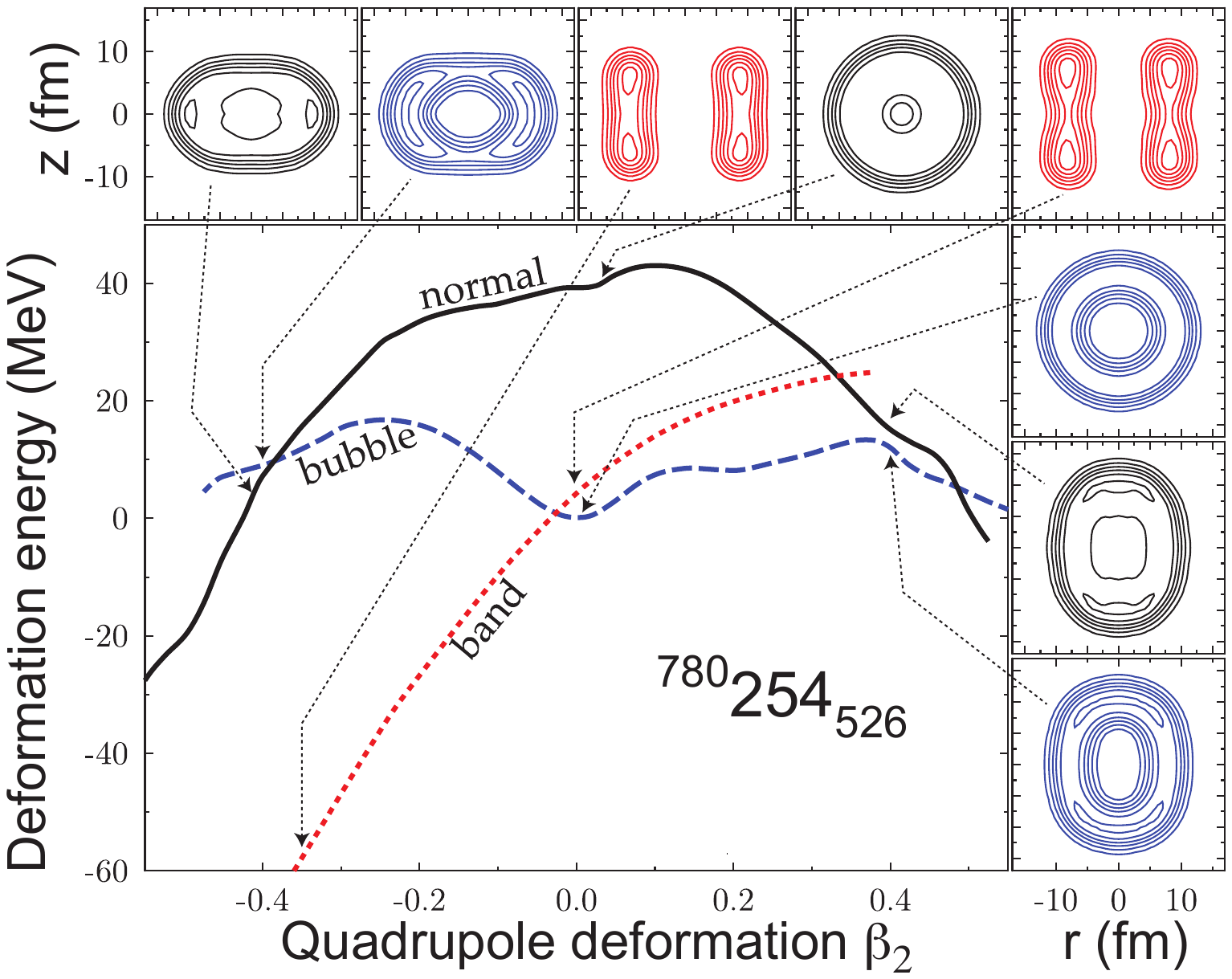}
\caption{Coexisting configurations associated with different density
distributions predicted by nuclear DFT for the
hypothetical superheavy nucleus $^{780}254_{526}$. Three topologies are considered: normal nuclear densities similar to those found in
stable nuclei, bubble nuclei distinguished by a substantial dip at the
center, and band configurations forming a thin band of nuclear matter
wound into a torus. The contour plots of the total densities are
given in the boxes.
(From Ref.\,\cite{Nazarewicz2002}).
}
\label{frustration}
\end{figure}

Competing short and long-range interactions lead to self-organization in domain patterns \cite{Seul1995}, \eg, in amorphous materials, glasses, or dilute magnets. In neutron stars, Coulomb frustration is responsible for the pasta phases in the inner neutron-star crust \cite{Pethick1995}. In atomic nuclei, frustration effects result in inhomogeneous phases produced in nuclear multifragmentation \cite{Chomaz2007}. In superheavy nuclei, the importance of Coulomb pressure increases with increasing system size. This favors a reduction of nuclear density in the nuclear interior and the appearance of exotic nuclear topologies such as bubbles or toroids \cite{Agbemava2015,Schuetrumpf2017,Jerabek2018,Afanasjev2018,Giuliani2019}. Figure~\ref{frustration} illustrates the competition between various configurations in the hypothetical superheavy nucleus $^{780}254_{526}$ \cite{Nazarewicz2002}, involving normal profiles (similar to densities of stable nuclei), bubbles which have a void at the center, and band-like toroids. In many cases, such exotic profiles are predicted to be energetically competitive. However, their stability in superheavy nuclei remains an open question. For that reason, the extent of the superheavy region is currently unknown \cite{smits2024}.

\subsection{Nuclear astrophysics}
Can superheavies be produced in Nature? This fundamental question is intrinsically related to determining which neutron-rich r-process nuclei are metastable against fission \cite{Guiliani2018,Giuliani2019}.

While superheavy nuclei produced in the r-process are all expected to be short-lived, they may impact the observed r-process abundances and electromagnetic transients produced by the radioactive decay of r-process nuclei in kilonovae \cite{Holmbeck2023b}. In another development, there have been searches for traces of superheavy elements in astrophysical data, including terrestrial ores, galactic cosmic rays, and meteorites \cite{Terakopan2017}. However, no definitive evidence of superheavy nuclei in nature has been found. A review covering the formation of the heaviest elements in the cosmos can be found in Ref.~\cite{Holmbeck2023a}.

\subsection{Atomic physics and quantum chemistry of superheavy elements}
Superheavy elements, because of their large atomic numbers and masses, represent a unique laboratory for relativistic effects and many-body phenomena due to the high density of electronic states \cite{Schwerdtfeger2020,Smits2023}.

\begin{figure}[tbh!]
\centering
\includegraphics[width=1.0\linewidth]{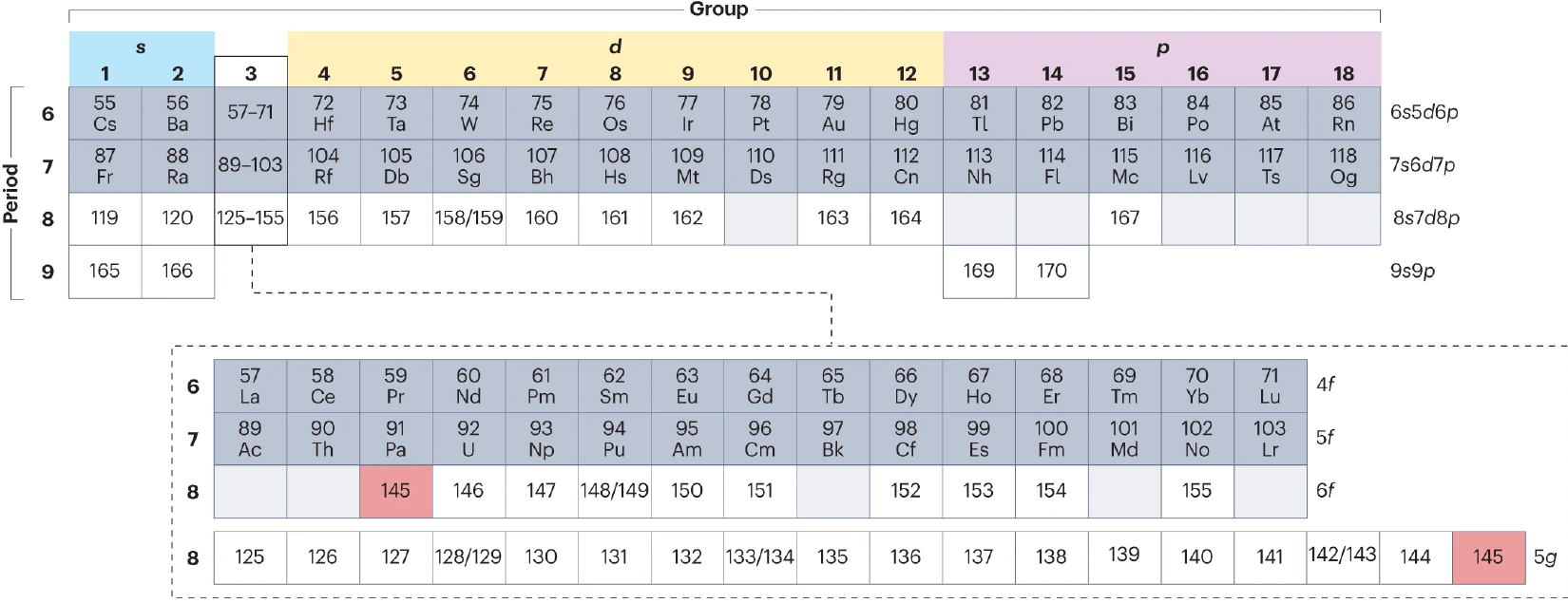}
\caption{An attempt to place the superheavy elements into the Periodic Table according to Dirac-Fock and configuration-interaction calculations \cite{Savelyev2023}. The unambiguous placement of elements $Z=121$, 122, 123, 124, and 168, the double placement of $Z=145$, and doubly occupied and vacant entries in the $8^{\text{th}}$ period highlight the breakdown of Periodic Table periodicity for superheavy elements.
(From Ref.\,\cite{Smits2023}).
}
\label{PT_shabaev}
\end{figure}

For physicists, the order of elements in the Periodic Table is fundamentally linked to the configuration of valence electrons. In superheavy elements, this ordering principle breaks down as different configurations lie energetically close, leading to ambiguity. Relativity, dense spectra, many-body correlation effects, and QED all play important roles that cannot be assessed by low-fidelity models. Figure~\ref{PT_shabaev} shows an attempt to place superheavy elements into the Periodic Table according to many-body calculations of Ref.~\cite{Savelyev2023}; it nicely illustrates the expected breakdown of Periodic Table periodicity.

While relativistic effects are known to impact properties of heavy elements such as lead \cite{Rajeev2011}, it is in the superheavy region that relativity dramatically changes basic atomic properties. Consider, \eg, Oganesson (Og) -- the heaviest known element ($Z=118$), completing the seventh period and group 18 of the noble gases (see Fig.~\ref{PT_shabaev}). Here, relativistic effects are expected to change the atomic size and shell structure \cite{Jerabek2018}, band structure (the solid form of Og is predicted to be a semiconductor) \cite{Mewes2019}, and state of matter (Og is expected to be a solid at ambient conditions with a melting point of about 325\,K \cite{Smits2020a}). Clearly, while placed in group 18 because of its valence electron structure, Og does not show similarities with its lighter noble gas homologues.

As discussed in Ref.\,\cite{Smits2023}, there are many connections between nuclear properties of superheavy nuclei and atomic properties of superheavy atoms. In both many-body systems, one finds very high densities of single-particle states, diffuse shell effects, and large electrostatic effects.

\subsection{Superheavy nuclei as open quantum systems}
All currently known superheavy nuclei are radioactive, and theory does not support the idea that stable superheavies would exist in nature \cite{smits2024}. Their main decay modes, fission and alpha decay, can be viewed as extreme examples of Coulomb frustration. The known superheavy nuclei are all proton-rich systems; they can decay by means of the forbidden channels of electron capture \cite{Ravlic2025}, but the branching ratios with respect to fission and alpha decay are small. The coexistence of several competing decay modes in superheavy nuclei makes them unique open quantum systems.

According to the Dirac equation, as the atomic number increases, the lowest $1s$ electron level enters the negative energy continuum \cite{Pomeranchuk1945,Smits2023}. In this supercritical region, the small (lower) component of the Dirac spinor becomes appreciable, and spontaneous electron-positron pair creation becomes possible as a hole in the Dirac sea escapes as a positron \cite{Gershtein1969,Pieper1969}. As discussed in Ref.~\cite{Smits2023}, in supercritical superheavy atoms one deals with a continuum space that contains metastable states (resonances) embedded in the non-resonant background. This situation resembles the behavior of quasi-particle resonances of the Hartree–Fock–Bogoliubov (HFB) equation in nuclear density functional theory (DFT). In fact, there are many similarities between the atomic single-particle Dirac problem and the nuclear HFB problem. This can be helpful when interpreting atomic phenomena in the presence of huge electric fields.

With the understanding of the relevance of SHN to nuclear physics and all these adjacent fields, we now focus on the mechanisms for production. 

\section{Understanding SHN production}
\label{sec:production}

\begin{figure}[tbh!]
    \centering
     \includegraphics[width=0.8\linewidth]{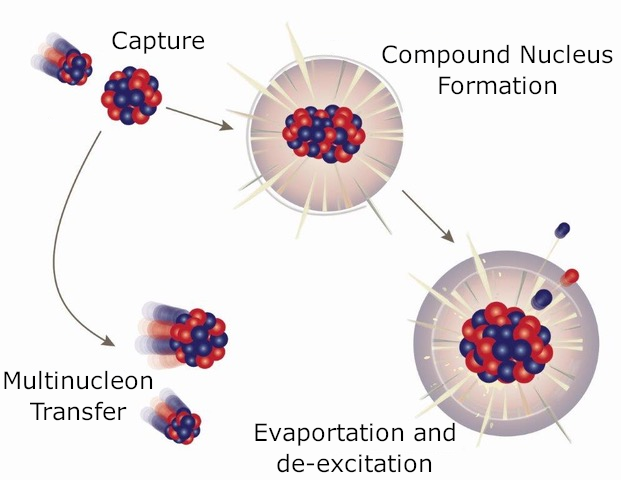}
    \caption{Cartoon illustrating the traditional three-phase model of superheavy nucleus formation: capture, compound nucleus formation, and
    evaporation.}
    \label{fig:cartoon}
\end{figure}

We first provide a birds eye view of the formation process.
Traditionally, the formation of a superheavy nucleus is assumed to consist of three distinct phases: i) capture, whereby the projectile overcomes the Coulomb repulsion, penetrates through the barrier and gets close to the target nucleus; ii) fusion, where nucleons from the two separate nuclei merge to form a compound nucleus; and iii) evaporation, when the compound nucleus decays, emitting particles and photons until it reaches the final product. Figure~\ref{fig:cartoon} provides a simple cartoon that illustrates this traditional three-stage view. Due to the separation of time scales between these stages, theory treats each phase independently using different models, which will be discussed in Sec.~\ref{sec:models}. 

The evaporation residue cross-section ($\sigma_{ER}$) is typically expressed as a product related to these three phases: $\sigma_{ER}(E,E_x) = \sum_{l} \sigma_{cap}(l,E) \times P_{CN}(l,E,E_x) \times W_{surv}(l,E_x)$, summing over angular momentum ($l$). $\sigma_{cap}$ is the capture cross-section, $P_{CN}$ is the probability of forming a compound nucleus once the projectile is captured by the target, and $W_{surv}$ is the survival probability against decay (evaporation phase).  These factors depend on the beam energy and/or the excitation energy of the compound nucleus, which are related via the Q-value for the reaction. 
Although the community recognizes that this three-phase picture is simplistic and that these phases overlap, it is currently not feasible to approach SHN production within a single formalism.

\begin{figure}[tbh!]
    \centering
     \includegraphics[width=1.0\linewidth]{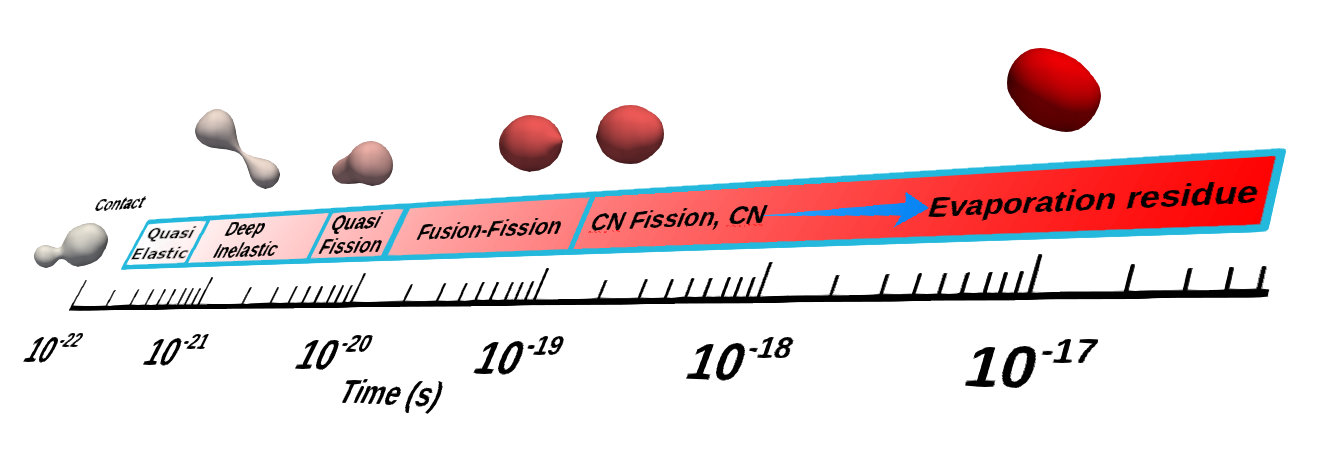}
    \caption{Typical timescales of heavy-ion collision stages at energies near the Coulomb barrier.}
    \label{fig:timescales}
\end{figure}

In low-energy heavy-ion reactions, many mechanisms can take place other than fusion. These mechanisms also involve a range of timescales and different sets of degrees of freedom, an example being multinucleon transfer (MNT), also shown in Fig.~\ref{fig:cartoon}.
See Figure~\ref{fig:timescales} for a schematic representation of the typical timescales of the different stages of near-barrier, heavy-ion collisions.
For these reasons, modeling such reactions presents significant conceptual and computational challenges.

The capture process is often understood as a barrier penetration, without much complexity in the reaction dynamics, although this picture has been shown to be too naive. The barrier penetration picture inherently contains ambiguity: What is the barrier, and how should its evolution be included as the nuclei approach each other? In reality, as the heavy-ion projectile approaches the target, many reaction channels open up, including breakup and transfer, which can greatly alter the saddle point in the collective space (see Sec.~\ref{sec:needscapture}), calling for a time-dependent approach. The intermediate phase of compound nuclei formation has received the most attention (many of the models covered in Sec.~\ref{sec:models} are focused on describing this phase), with a diverse set of microscopic and macroscopic theories developed, most of which are rooted at some level in phenomenology. The final evaporation phase is often modeled as a survival probability using a statistical Hauser-Feshbach approach and is an area where there has been less development, as we discuss in Sec.~\ref{sec:needssurvival}.

For decades, the experimental community has been invested in this field. In recent decades, experimental efforts to produce new superheavy elements have utilized cold and hot fusion reactions, often involving \nuc{48}{Ca} beams on actinide targets. Along these lines, the discovery in Dubna of the heaviest element in the periodic table ($Z=118$ Oganesson \cite{Oganessian15}) brought great excitement to the field. But it has taken nearly a decade for the next step to be completed, namely the proof of principle by Berkeley that $^{50}$Ti can be an effective projectile for producing the next superheavy element ($Z=120$) \cite{gates2024}. Alternatively, laboratories have been using heavier projectiles (such as Xe) to induce multi-nucleon transfer on heavy targets. Following earlier work at Texas A\&M \cite{Natowitz08,Barbui09} and theory predictions by \cite{Zagrebaev2007}, a recent multi-nucleon transfer campaign with KISS in RIKEN has performed spectroscopy on superheavies and unveiled large matter radii for neutron-rich platinum isotopes \cite{Hirayama2022}. To better understand the complexity of the process, the Legnaro experiment measured a wide range of products on the path to fusion of $^{48}$Ca+$^{208}$Pb, exposing the diverse reaction channels occurring simultaneously with fusion \cite{Cook2023}. 

Experiments with SHN typically require very long runtimes due to low statistics. To form a compound nucleus, the beam energy needs to be precisely within the right window: too little energy and the projectile will not be able to overcome the barrier; too much energy and the resulting compound state has insufficient time to equilibrate and falls apart. Each phase presents experimental challenges and opportunities requiring strong theory-experiment interplay.  From an experimental perspective, theory predictions are vital for identifying the most scientifically significant measurements, aiding interpretation, and refining predictions for new element searches. Furthermore, experiments to produce new SHN rely critically on model estimates so the experimental conditions can be optimized. In the next section, we briefly review a number of models used in this context, along with some opportunities for improvement.


\section{Theoretical Frameworks for Superheavy Nuclei Production}
\label{sec:models}

The synthesis of SHN in heavy-ion collisions involves a complex sequence of events: initial capture, dynamic evolution towards a compact configuration, and eventual formation and decay of a compound nucleus. Ideally, a single, self-consistent theoretical framework could describe this entire process, encompassing the structure of the colliding nuclei, the interplay of collective and intrinsic degrees of freedom, and the transition from rapid, non-equilibrium dynamics to a statistically equilibrated system. However, as mentioned before, due to the vastly different time scales involved and the inherent complexity of the quantum many-body problem, such a unified, real-time description is not currently feasible.

A central challenge in modeling SHN fusion is describing the transition between rapid, non-equilibrium (diabatic) processes and slower, gradual (adiabatic) motion towards the compound nucleus. In mean-field descriptions, this involves nucleons potentially not following the lowest energy orbitals at level crossings along the collective path, leading to distinct diabatic and adiabatic potential energy surfaces (PES) \cite{Nazarewicz1993}. While this transition regime is not directly observable experimentally given the extremely brief timescales, insights from quasifission characteristics (mass splits, total kinetic energies, scattering angles, rotation timescales) and studies of hindrance mechanisms \cite{Hinde2021, Itkis2022} provide crucial guidance for theoretical model development.

Given these complexities and the limitations of single approaches, the field employs a variety of theoretical models, each tailored to describe specific aspects or stages of the reaction. These models often rely on different approximations. To provide a more comprehensive picture and bridge the gaps between models, hybrid approaches that combine different theoretical frameworks are increasingly utilized. Despite potential formal inconsistencies, these hybrid models can effectively reproduce experimental phenomena and offer valuable predictive power, particularly when combined with rigorous uncertainty quantification.

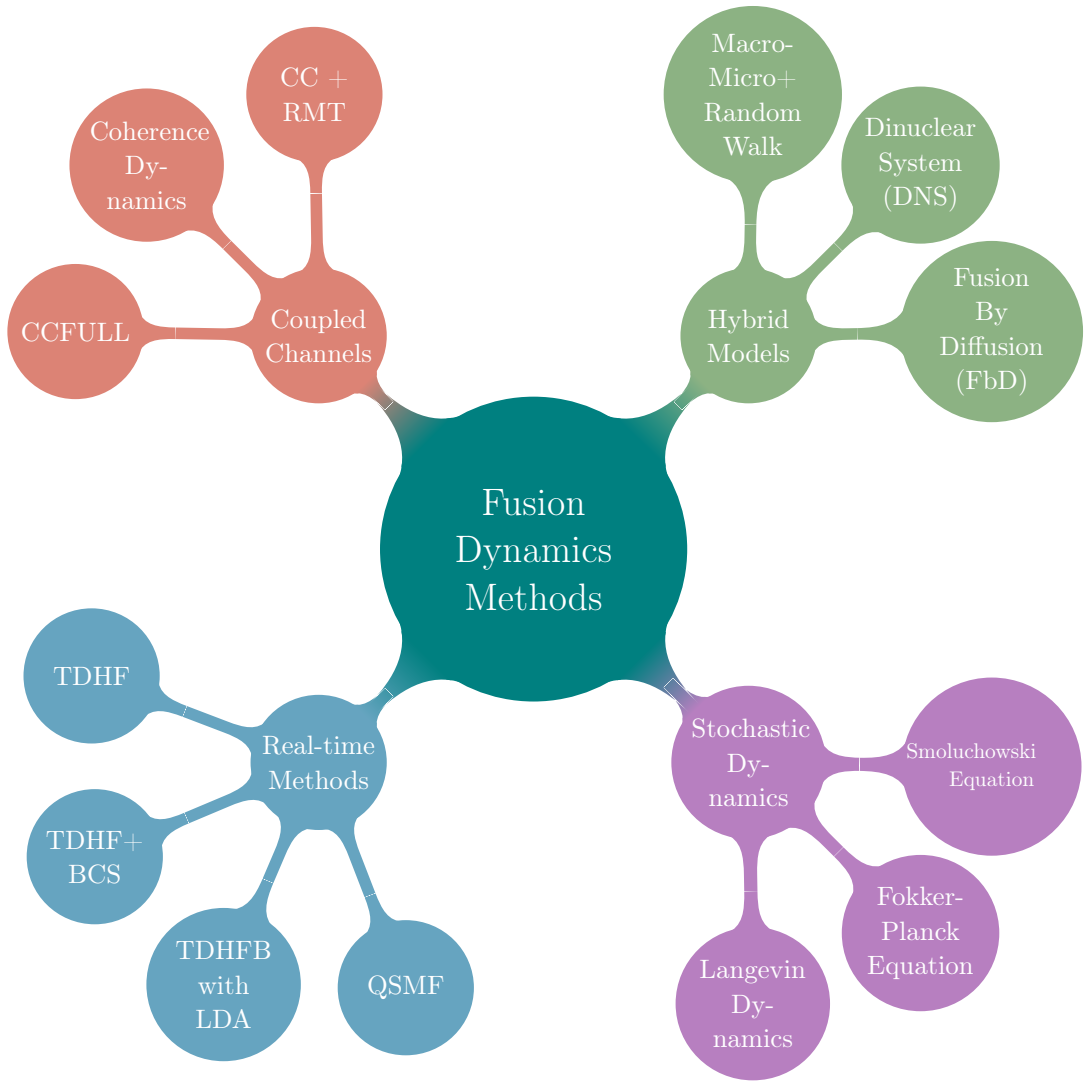
\begin{figure}[tbh!]
    \centering
\begin{tikzpicture}[mindmap, grow cyclic, every node/.style=concept, concept color=teal, text=white, level 1/.append style={level distance=4cm, sibling angle=90}, level 2/.append style={level distance=3.2cm, sibling angle=44}]
  \node[concept] {Fusion Dynamics Methods}
    child[concept color=myOrange!60] {node[level 2] {Real-time Methods}
      child { node {TDHF} }
      child { node {TDHF+\\BCS} }
      child { node {TDHFB with LDA} }
      child { node {QSMF} }
    }
    child[concept color=myPlum!60] {node[level 2] {Stochastic Dynamics}
      child { node {Langevin Dynamics} }
      child { node {Fokker-Planck Equation} }
      child { node[minimum size=2.9cm,scale=0.8] {\hspace{-19pt}Smoluchowski Equation} }
    }
    child[concept color=myGreen!60] { node[level 2] {Hybrid Models}
      child { node {Fusion By Diffusion (FbD)} }
      child { node {Dinuclear System (DNS)} }
      child { node {Macro-Micro+\\Random Walk} }
    }
    child[concept color=myRed!60] { node[level 2] {Coupled Channels}
      child { node {CC + RMT} }
      child { node {Coherence Dynamics} }
      child { node {CCFULL} }
    };
\end{tikzpicture}
    \caption{Visual overview of theoretical methods used in describing fusion dynamics towards superheavy nuclei.}
    \label{fig:mindmap}
\end{figure}

This section briefly overviews theoretical approaches commonly used for heavy-ion fusion leading to SHN, focusing on their applications and highlighting opportunities for improvements and advancements relevant to future experiments. Figure~\ref{fig:mindmap} provides a visual overview of the methods here covered.

\subsection{Time-Dependent Hartree-Fock (TDHF)}
TDHF describes nuclear dynamics by evolving single particles in a self-consistent mean field. In the context of SHN production, it is used to calculate the initial dynamics of the colliding system and to extract properties related to fusion probabilities.
TDHF has been used extensively to study the effects of projectile/target orientation and nuclear shell structure on fusion and quasifission outcomes \cite{umar2010, simenel2012_influence, umar2014, umar2015, wakhle2014, umar2016,sekizawa2016}. TDHF simulations also provide insights into multinucleon transfer and the role of nuclear structure in dynamical evolution~\cite{sekizawa2013, sekizawa2017, guo2018_role, guo2018_influence,godbey2019,godbey2020}. For recent reviews of TDHF applications in heavy-ion dynamics, see Refs.~\cite{simenel2012_nuclear, simenel2018,sekizawa2019tdhf,godbey2020quasifission}.

The primary phenomenology in TDHF lies in the choice of an effective interaction or energy density functional. While results are qualitatively similar across different density functionals used in TDHF, the basic formulation lacks pairing correlations which can influence the diabatic-to-adiabatic transition. The mean-field approximation also omits explicit two-body collisions and quantum fluctuations, limiting its description of system evolution and classically forbidden motion. This exclusion of pairing, in particular, might overestimate the transfer and equilibration timescales.

Integrating pairing requires extensions like Time-Dependent Hartree-Fock-Bogoliubov (TDHFB) and TDHF + BCS (TDBCS). While more efficient, TDBCS is an approximation to TDHFB that can lead to unphysical results, particularly for particle emission due to violation of the continuity equation \cite{Scamps2012}. Dynamic models with pairing highlight how pairing couplings differ during level crossings \cite{blocki1976}. Nuclear density functional theory studies, including mean-field and pairing correlations \cite{reinhard1997}, and recent TDHFB simulations emphasize pairing's role in dissipative processes. Time-dependent density functional theory (TDDFT) with the Superfluid Local Density Approximation (SLDA/TDSLDA) includes the dynamics of the pairing field, where phase differences can create solitonic structures, potentially introducing additional barriers to fusion~\cite{bulgac2013, magierski2016}.

\subsection{Stochastic Dynamics}

At energies where SHN are produced in fusion-fission-like events, heavy-ion collisions exhibit dissipative behavior, converting collective energy into intrinsic excitation and raising the system's temperature.
Describing these non-equilibrium collisions using stochastic dynamics models relies on a number of assumptions such as a large number of nucleons, a thin nuclear surface, low temperature relative to Fermi energy, long mean free path (Knudsen gas), and semiclassical collective velocities \cite{Feldmeier1987}.
A key conceptual difficulty is identifying appropriate macroscopic variables and defining the concept of a ``heat bath''. These models often employ bulk nuclear properties and concepts from non-equilibrium statistical physics, although they typically do not explicitly incorporate detailed nuclear structure effects.
The theoretical description of these processes evolved from the challenge of understanding fusion hindrance in massive systems. Overcoming the static Coulomb barrier is not a sufficient condition for fusion, as strong dissipative forces can drive the system towards re-separation through quasifission.
This phenomenon was first addressed phenomenologically by Świątecki’s ``extra-push'' model, which posited that an additional kinetic energy above the barrier is required to surmount this dynamical hindrance \cite{Swiatecki1981, Swiatecki1982}.
This concept was supported by experiments showing rapid dynamical deformations \cite{Sann1981, Bock1982, Westmeier1982}, and its microscopic origin was initially connected to one-body dissipation, as formalized in the wall-and-window model \cite{Blocki1978}. However, the precise nature of this hindrance and the underlying concept of an `additional inner barrier' remain a key, open question. Recent theoretical investigations using Time-Dependent Density Functional Theory (TDDFT) have revealed that nuclear superfluidity can also introduce a significant hindrance, creating an effective energy barrier conceptually similar to the classical extra-push \cite{Magierski2017}.
This pairing-induced effect suggests that the observed hindrance is not solely due to macroscopic dissipation. A fundamental challenge thus persists: is the total hindrance simply the integrated effect of all competing non-equilibrium channels, such as quasifission, or is there an additional, intrinsic dynamical mechanism that obstructs the system's trajectory toward the compound nucleus?
To model dissipative and stochastic phenomena, theoretical frameworks were developed, building upon the foundational work of H. A. Kramers \cite{Kramers1940}. Early approaches utilized the Fokker-Planck equation (FPE) to describe dissipation and fluctuations in deep-inelastic collisions \cite{nirenberg1974fokker, ngo1977multidimensional, berlanger1978multidimensional, agassi1978multidimensional, yadav1982multidimensional, frobrich1983multidimensional, bohne1983multidimensional, feldmeier1985multidimensional}. Subsequently, the implementation of the Langevin equation, which respects the fluctuation-dissipation theorem, allowed for a probabilistic treatment of competing channels \cite{Feldmeier1987, Schmidt1991, Przystupa1994, Abe1996, Frobrich1998, Abe2000, Shen2002, Boilley2003}. The power of this approach was substantially advanced by Aguiar et al., who provided a unified description of outcomes from deep-inelastic scattering to complete fusion \cite{Aguiar1989, Aguiar1990}, and by Wada et al., who demonstrated that including one-body dissipation in Langevin simulations was essential for reproducing experimental data \cite{Wada1993}. This evolution has established multidimensional Langevin models as the standard tool for predicting the probabilities of superheavy element formation \cite{Zagrebaev2015, Aritomo2012}.
The Langevin formalism encompasses friction and Langevin forces, which describe the coupling between collective and intrinsic degrees of freedom, driving fluctuations in heat and energy. Significant challenges remain, particularly in the microscopic determination of the friction and mass tensors. The selection of collective variables is also crucial but often arbitrary, though methods like the Fourier-over-spheroid (FoS) parametrization show promise \cite{Pomorski2023}. Furthermore, the initial geometry, such as the orientation of deformed nuclei in the entrance channel, must be considered \cite{Amano2024, Zagrebaev2007}. Despite these challenges, Langevin simulations effectively reproduce a wide range of experimental data, providing valuable insights into measurable quantities and reaction time scales.
Under the assumption of strongly overdamped dynamics, where inertial effects are negligible, the Langevin equation simplifies to the Smoluchowski equation. The premise of strongly damped motion is supported by arguments from the wall-and-window model \cite{Blocki1978, Sierk2017}, transport models \cite{Ivanyuk1999, Hofmann2001}, and more recent microscopic arguments and simulations \cite{Sierk1980, Schutte1980, Tanimura2015, Bulgac2019}.

\subsection{Fusion by Diffusion (FBD) Model}

 The Fusion-by-Diffusion (FBD) model, which describes the formation of a compound nucleus as a diffusion process over a barrier, is formulated using the Smoluchowski equation. This approach builds upon a strong theoretical heritage. Foundational work by Weidenmüller and Zhang applied a stationary diffusion framework to describe nuclear fission over a multidimensional potential barrier \cite{Weidenmuller1984}. Subsequently, Aritomo et al. employed a similar dynamical approach to investigate the formation of superheavy elements, highlighting the importance of dissipative effects \cite{Aritomo1997}. These pioneering studies established the validity of diffusion-based models for describing complex nuclear dynamics, paving the way for the development of the FBD model. The FBD model \cite{FBD-Acta,FBD-05,FBD-11,PRC-hot,Hagino2018FBD}, , which primarily describes the middle stage of fusion as an overdamped process that occurs after overcoming the Coulomb barrier. It assumes that the remaining kinetic energy is converted into internal excitation. The model simplifies the process to one effective dimension where the system must overcome a potential barrier to fuse. In the infinite time limit, the Smoluchowski equation has an analytical solution.

Fusion in FBD requires overcoming a saddle point via thermal fluctuations in shape degrees of freedom. The energy threshold for fusion is the difference between the saddle point energy and the system's energy at the ``injection point," corrected for rotational energies. The model posits that after passing the entrance channel barrier, the neck expands rapidly, placing the system at an ``injection point" in the asymmetric fusion-fission valley of the compound nucleus PES. This point marks the start of diffusion.

The advantage of FBD is its analytical solution for subbarrier and above-barrier fusion, allowing for the calculation of the compound nucleus formation probability across energies. It applies to both cold (low excitation) and hot (high excitation) synthesis.

Challenges include guessing the injection point (a key parameter adjusted from known reactions, which has been addressed in Hybrid models as described in Sec.~\ref{Sec:HybridModels}), its inherent one-dimensional nature limiting capture of multi-dimensional dynamics, the lack of explicit quantum shell effects in the dynamics, and the reliance on a conditional fusion saddle point specific to the model.

\subsection{Dinuclear System (DNS) Model}
The DNS model focuses on the mass asymmetry and relative distance between colliding nuclei \cite{zubov2003}. After capture into a potential pocket, relative kinetic energy converts to potential and excitation energy \cite{adamian2008}. The DNS evolves over time via diffusion along the mass asymmetry coordinate \cite{adamian2012}.

DNS uses repulsive diabatic potentials beyond the touching point, preventing merging along the internuclear distance, based on the idea that nuclei do not easily integrate \cite{diaz-torres2000, antonenko1993}. In heavier systems, nuclei remain in a touching configuration, forming a DNS, with nucleon exchange occurring until the smaller nucleus is absorbed, forming a compound nucleus \cite{antonenko1995, adamian1997, adamian1997b, adamian1998, giardina2000, volkov2004, zhao2008}. This aligns with observations in heavy and superheavy nuclei production \cite{adamian2012}.

In this model, the fusion probability is defined as the likelihood of the DNS overcoming an inner fusion barrier along the mass asymmetry coordinate. The DNS can decay by overcoming the quasifission barrier or transitioning to symmetric configurations more prone to decay. Recent DNS predictions suggest optimal reactions and estimate evaporation residue cross-sections for elements 119 and 120 \cite{adamian2022}.

Both FBD and DNS models describe  capture, fusion and  survival probabilities, and therefore are able to predict the total evaporation cross-sections. However, both simplify the fusion process to one dimension and rely on a conditional fusion saddle point.

\subsection{Coupled Channels (CC) Methods}
Coupled channel methods are particularly suitable for energies around and below the Coulomb barrier, where quantum tunneling and coupling to intrinsic nuclear motion is critical. These methods consider the coupling between the relative motion of two nuclei and their internal excitations (vibrations, rotations). CC calculations were initially fitted to scattering data; for fusion, incoming wave boundary conditions and a short-range imaginary potential are used to account for capture. CC methods require empirical information on projectile/target degrees of freedom. 

The coupled channel method is also used to produce fusion barrier distributions, crucial for SHN synthesis. For example, the experimental barrier distributions for $^{48}$Ca + $^{248}$Cm \cite{tanaka2018} have been compared with the CC calculations obtained using CCFULL \cite{Hagino1999} and including rotational and octupole excitations as well as one-neutron transfer. A major challenge is incorporating multinucleon transfer processes fully quantum mechanically, consistent with inelastic channels and accurate Q-value distributions. Transfer channels can be included when using FRESCO \cite{fresco} where   fusion is indirectly extracted through  the absorption cross section (see \cite{Jia2025} for a recent application).

\subsection{Coherence Dynamics}
A dynamical coupled-channels density matrix method based on open quantum systems theory has been used to describe low-energy nuclear fusion \cite{LeeDiazTorres2022}. Combined with an energy projection approach, it calculates energy-resolved fusion probabilities, showing excellent agreement with stationary Schrödinger dynamics. Studies using this framework can calculate entropy, energy dissipation, and coherence, demonstrating that quantum decoherence does not impact fusion probabilities. Recently, this approach has been used to determine friction, incorporating a phenomenological nuclear friction form factor and coherent coupled-channels effects \cite{lee2024}.

\subsection{Hybrid Models}\label{Sec:HybridModels}
The complex interplay of quantum and dissipative effects near the Coulomb barrier also motivates the development of hybrid approaches, some of which are explored in this section.\\

\noindent{\it CC + Random Matrix Theory (RMT)}: This approach addresses phenomena not explained by simple potential models. A random-matrix model for complex excitations \cite{Ko1976,Agassi1978a,Barrett1978,Agassi1977,Ko1979,Agassi1978b} is combined with CCFULL \cite{Yusa2013,Yusa2013b,Piasecki2019,Piasecki2020} to study quasi-elastic barrier distributions.\\

\noindent{\it TDHF + Langevin}: Quasifission is a main competitor to SHN synthesis. TDHF describes the entrance channel and quasifission dynamics on average. This hybrid approach extracts post-capture configurations from TDHF and uses them as initial conditions for a Langevin-type description of compound nucleus formation, often within a fusion-by-diffusion framework \cite{Sekizawa:2019bhj}. This was applied to Z=120 reactions, clarifying the role of $^{48}$Ca magicity \cite{Sekizawa:2019bhj, Sun:2022aad, Sun:2023zri}.\\

\noindent{\it Quantum Stochastic Mean-Field (QSMF)}: This quantal transport approach calculates isotope production cross-sections in MNT reactions \cite{Kayaalp, Yilmaz2020, Ayik2017, Ayik2018, Yilmaz2018, Ayik2019, Sekizawa2020, Ayik2020, Ayik2020b, Ayik2023, Ayik2019b, Ayik2023b, Arik2023}. It provides insights into production distribution, interaction, and nucleon exchange dynamics. Contributions from statistical evaporation are added (e.g., using GEMINI++ \cite{Yilmaz2020,Sekizawa2020}). Coupled differential equations for variances account for diffusion and curvature from the PES, with diffusion coefficients derived from TDHF single-particle states. This allows QSMF to reproduce the measured width of  fragment distributions, unlike average mean-field theories.\\

\noindent{\it  Dynamical Cluster-Decay Model (DCDM)}: Another theoretical framework used to describe SHN production is the Dynamical Cluster-Decay Model (DCDM) \cite{Kumar2013, Sawhney2015}. This model treats the decay of the compound nucleus as a dynamic process in which preformed clusters tunnel through the interaction barrier. The model is described in terms of collective coordinates, such as mass and charge asymmetry, and has been widely applied to calculate fusion cross-sections, evaporation residue yields, and decay properties in the superheavy region.
\\

\noindent{\it Microscopic-Macroscopic (mic-mac) + Random Walk}: This hybrid method uses the random walk approach on energy surfaces calculated with the mic-mac method. The mic-mac model divides the total energy into a smooth macroscopic part (e.g., liquid-drop) and a microscopic part consisting of  shell and pairing corrections \cite{Brack1972, baran2005,ward2017,verriere2021}. It calculates deformation energy surfaces as functions of collective variables (shape parameters). The random walk method explores these surfaces, with steps weighted by the density of states, allowing the system to navigate different shapes, including asymmetric ones. Rotational energy is included. This approach accurately reproduces nuclear masses and fission barriers for heavy nuclei and describes structural properties across a wide range of $Z$ and $N$~\cite{randrup2011,randrup2011a,mumpower2020}. The random walk approach typically applies only above the fusion barrier. It relies on the assumption of a Fermi gas for highly excited systems, and it is limited to the deformation space explored.\\

While the models used are many and varied, there are specific questions, pertaining to each of the three phases, that remain open. Model improvements and new theory developments are needed to address these as discussed in the next section.

\section{Theory needs and future developments}
\label{sec:needs}

\subsection{Capture}
\label{sec:needscapture}

SHN synthesis begins with the capture process, leading to a structured barrier distribution through induced excitations and nucleon transfer \cite{Hagino2022}. While capture is relatively well understood compared to compound nucleus formation and survival, reducing phenomenological assumptions is desirable, especially for reactions reaching unexplored regions of the heaviest elements. Combining knowledge from microscopic calculations with the CC method is a promising direction.

Microscopic approaches like TDHF(B) can provide inputs for CC calculations, such as excitation energies from linear response TDHF or bare nucleus-nucleus potentials from density-constrained HF(B) \cite{Simenel2013}. This helps determine CC phenomenological parameters microscopically. CC calculations using excitation energies and couplings from beyond-mean-field covariant DFT studies can include anharmonicity effects, like the coupling between quadrupole and octupole phonons shown to be important for $^{16}$O+$^{208}$Pb fusion \cite{Hagino2015,Yao2016}.

Nucleon transfer also affects capture. While standard CC includes few-nucleon transfer channels \cite{Esbensen1998}, large-scale transfer can alter Q-values and Coulomb barriers, significantly impacting fusion~\cite{godbey2017}. A fundamental question is \textit{``What is actually fusing?''} since transfer can change projectile/target identity during the collision. Quantifying this effect requires extracting transfer probabilities from TDHF wave functions \textit{during} the collision using particle-number projection \cite{simenel2010}, which is feasible when a separation plane exists.

Significant theoretical effort is needed to clarify the effects of entrance-channel transfer on barrier and excitation-energy distributions. This is crucial for SHN synthesis. The detailed analysis of $^{40}$Ca+$^{208}$Pb collisions in Ref.~\cite{Cook2023} showed $\sim$20--30 populated nuclides with excitation energies up to 60 MeV near or below the barrier, highlighting the need for a consistent microscopic description of the diverse fusing nuclei identities. Describing heavy-ion deep-inelastic collisions (DIC) and fragment distributions is related. While TDHF with particle-number projection estimates few-nucleon transfers correctly, it is unable to reproduce the data when the number of transferred nucleons is larger than a couple. Extended approaches like TDRPA \cite{simenel2012} or QSMF \cite{Kayaalp} improve transfer probability descriptions. Ideally, a ``multi-configuration TDHF(B)'' could describe a variety of  observables, such as transfer cross sections, angular momentum distributions, Q-values, excitations, and barrier distributions, providing information on the transition to a thermalized compound nucleus on the path to fusion \cite{simenel2012_influence}.

\subsection{Compound nucleus formation}
\label{sec:needscompound}

Following capture, the captured system's energy must dissipate into the compound nucleus's internal degrees of freedom. Crucially, the system may reseparate via quasifission before forming the compound nucleus, preventing the desired SHN production. Theoretical efforts must focus on the timescales and mechanisms of dissipation and reseparation, to accurately calculate quasifission and fusion probabilities ($P_{CN}$).

TDHF suggests at least two equilibration timescales: kinetic energy and angular momentum equilibrate rapidly ($\sim 10^{-21}$ s), as does the $N/Z$ ratio. Full mass equilibration takes longer ($\sim 10^{-20}$ s), comparable to rotation timescales relevant to experimental quasifission measurements \cite{simenel2020}.

Mic-mac models offer a computationally advantageous alternative to fully microscopic models like TDHF(B), demonstrated in fission yield descriptions. However, quasifission's complex shapes require models capturing many moments. Comparing mic-mac and microscopic model predictions could assess validity and align results with experiments.

Understanding the compound nucleus temperature dynamics, and its link to excitation energy, remains challenging. TDHF and TDHFB account for one-body dissipation within the mean-field approximation. A realistic temperature and dissipation picture may require correlations beyond the mean field. Mic-mac models incorporate temperature into fission calculations by affecting the PES and random-walk parameters, but may not fully describe the post-capture phase before the system becomes overdamped. In general, a deeper understanding of finite temperature effects will further refine our understanding of fusion and fission dynamics.

This raises the question of how time-dependent paths in microscopic models compare to Markov-chain paths in mic-mac approaches. Both apply a smooth exploration of deformation space, but a clear link between time and random walk steps is needed for a coherent discussion of a given path's significance. Microscopic models are better suited for describing particle (especially neutron) evaporation during the CN evolution, which is crucial for neutron-rich SHN production.

Understanding what determines the competition between fusion and quasifission is fundamental: Is it CN properties alone, capture configurations, $N/Z$ ratio, or shell effects? Shell effects are a primary driving force for quasifission, potentially hindering CN formation by halting equilibration. Studying the connection between shell effects in quasifission and fission could significantly increase our understanding of SHN production \cite{simenel2021}. Preliminary theoretical studies imply a strong correlation, but systematic theoretical and experimental studies are needed. The question is not {\bf if} shell effects exist, but {\bf how strong} they are for a given system. TDHF studies show shell effects vary by system, and deformed shells tend to dominate over spherical ones \cite{oberacker2014b,godbey2019}.

Experimental probes clearly show the need for including the effect of closed shells on quasifission distributions. Past efforts suggested $^{208}$Pb shells were important, though later interpretations could suggest ternary fission contributions (where the neck rupture forms a third fragment). Signatures of deformed shells ($Z=54$--$56$) in quasifission are predicted by a number of TDHF simulations, for different systems and different effective interactions~\cite{godbey2019,simenel2021,li2022impact}, though these predicted fragment distributions have not been clearly seen experimentally. A theoretical prediction of inverse quasifission with a strong deformed or spherical shell gap signature might offer a less ambiguous comparison as its outcome does not overlap with symmetric fusion-fission yields~\cite{simenel2010iqf}. A measurement of Gd$+$W reactions noted an inverse quasifission channel (quasifission with large mass asymmetry) attributed exclusively to the spherical shell gaps in \nuc{208}{Pb}. It should be noted that the light fragment in that case would have $Z=56$, suggesting that it is a combination of the two effects, the deformation effect from the light fragment and the spherical effect from the heavy fragement,  that leads to an enhancement in the cross section of that channel.
Additionally, ideal measurements would also include mass and TKE distributions with TDHF and mic-mac predictions for quasifission outcomes to permit a systematic comparison between models.

For systems with open shells in fragments, pairing becomes important. Opening a shell introduces a pairing phase difference between projectile and target as a new degree of freedom, potentially altering the Coulomb barrier \cite{magierski2017novel, magierski2022pairing}. This pairing phase difference raises new questions: How does pairing behave in asymmetric collisions? How does the pairing phase difference alter transfer? How does pairing change CN energy dissipation, affecting survival probability \cite{makowski2023}? 

FBD and random-walk models assume overdamped shape evolution. Defining when this assumption is valid post-capture (determining the initial shape configuration/injection point) is important. A recent study \cite{albertsson24:a} (building on \cite{albertsson21:a}) assumes initial radial drift with one-body friction \cite{Blocki1978} until the radial speed drops. Then the system diffusives through random walks in a 5D PES, guided by shape-/energy-dependent level densities. Results with these assumptions agree reasonably well with data for $^{208}$Pb targets but should be tested for other cases.

\subsection{Evaporation phase}
\label{sec:needssurvival}

Heated heavy nuclei are highly prone to fission in addition to particle evaporation. To produce SHN, the compound nucleus needs to evaporate particles (neutrons, charged light particles) without fissioning. This is usually expressed in terms of a final survival probability ($W_{surv}$). This quantity is critical for synthesizing nuclei in hot reactions where intense heating leads to multiple neutron emissions. Accurate estimation of $W_{surv}$ is a key ingredient in the planning of future hot synthesis experiments aiming at discovering new elements.

\begin{figure}[h!]
 \includegraphics[width=0.49\textwidth]{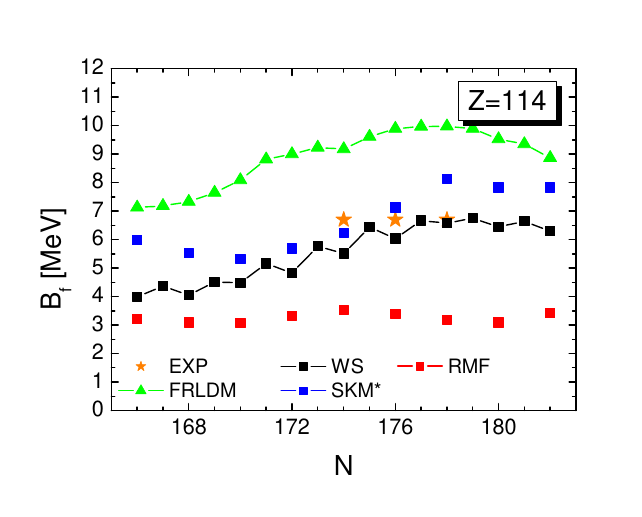}
  \includegraphics[width=0.49\textwidth]{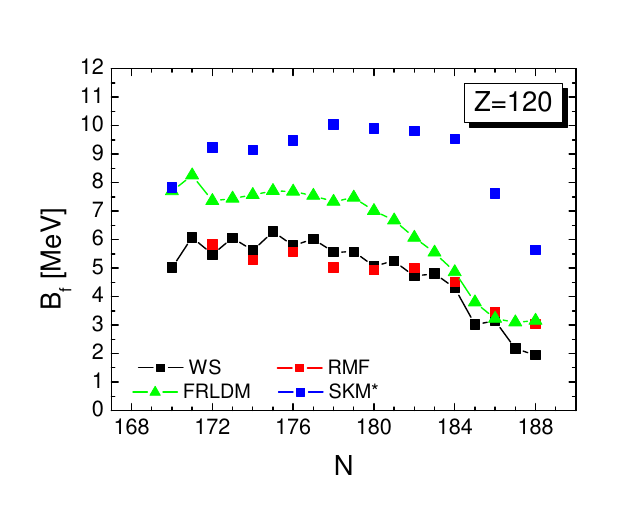}
 \caption{
 Fission barriers predicted by various models for $Z=114$ (left panel) and for $Z=120$ (right panel) isotopes: black - WS model, green – FRLDM \cite{FRLDM}, blue – SkM* \cite{SKM}, red – RMF with NL3 parametrization \cite{RMF}. Experimental data taken from \cite{ITKIS}. Figure taken from \cite{Jachimowicz2017a}.
 }
 \label{fig:barriers} 
\end{figure}

$W_{surv}(E_x, J)$ measures the balance between fission decay width ($\Gamma_f$) and particle evaporation width ($\Gamma_n, \Gamma_p, \Gamma_\alpha$) from the excited CN. For SHN at relevant excitation energies, neutron emission often dominates over $\gamma$-rays and charged particles (due to the Coulomb barrier). The survival probability after $x$-neutron sequential emissions ($W_{xn}$) is proportional to a product of ratios of these partial widths $\Gamma_n/(\Gamma_f + \Gamma_n)$ from each neutron emission. Usually, the Weisskopf formula is used for $\Gamma_n$ and transition-state theory is used for $\Gamma_f$. The ratio $\Gamma_n/\Gamma_f$ is exponentially sensitive to the difference between the square-roots of the fission barrier ($B_f$) and neutron separation energy ($B_n$). A 1 MeV increase in $B_f$ can enhance cross-sections by 2--3 orders of magnitude, highlighting the need for accurate nuclear inputs, especially in multi-stage cascades. 
Masses, for instance, are vital pieces of nuclear data, particularly for systems far from stability. They are needed to compute the Q-values for the reactions, an ingredient necessary to relate the beam energy to the excitation energy of the populated compound nucleus state.
Furthermore, mass studies must be comprehensive (including even-even, even-odd, and odd-odd nuclei), given that successive neutron emission will traverse several isotopes in a chain. Small uncertainties in the masses can produce large uncertainties in the resulting $W_{surv}$.
Accurately estimating $W_{surv}$ for multi-$xn$ channels also requires understanding fission barriers for whole isotopic chains including odd nuclei, emphasizing the need for systematic calculations.  Figure \ref{fig:barriers} shows predicted barriers for $Z=114$ and $Z=120$ as a function of neutron number.
Large discrepancies between various models can be seen.

Figure \ref{fig:barriers}(left) contains the results for Flerovium's (Fl, $Z=114$) and includes experimental data when available. The self-consistent Skyrme HFB calculations with the SkM* functional \cite{SKM} are able to reproduce the data, but others are not. For example, FRLDM \cite{FRLDM} overestimates the difference  and RMF \cite{RMF} (before tuning \cite{RMF2015}) underestimates the barriers, leading to problematic predictions regarding known millisecond half-lives. For $Z=120$ (for which there is no data), models vary significantly, with Skyrme SkM* predicting the highest barriers likely due to a proton magic gap. Self-consistent theories predict barriers around 10 MeV which would strongly hinder fission and therefore suggests larger SHN production yields---a result at odds with cross section estimates from unsuccessful synthesis attempts.


Besides $B_f$, nuclear level densities are crucial. $B_f$ sets integral limits for the density of states above it, directly impacting decay probability. Reliably determining nuclear level densities remains a significant challenge despite its known importance. The calculation of nuclear level densities from many-body calculation is done by counting states, and involves determining eigenvalues and degeneracies of the nuclear Hamiltonian.
It is impossible to do at high excitation energy, requiring statistical methods. Various combinatorial and microscopic methods (e.g., based on RMF \cite{pomorska1, pomorska2}, Yukawa potential \cite{pomorska3}, shell model Monte Carlo \cite{Alhassid}, moment method \cite{Senkov}, or the Gogny interaction \cite{hilaire2012}) exist, often requiring approximations to  superfluidity phenomena and collective enhancements.

Given the many loose ends discussed in this section, we found that a systematic assessment of all inputs for $W_{surv}$ calculations is needed, including parameterizations for odd nuclei, their impact on masses, deformations, saddle points, and nuclear level densities. Developing better global models with quantified uncertainties for these inputs is necessary. Comparing predictions from Microscopic-Macroscopic and DFT-based approaches for $B_f$ and nuclear level densities in SHN is also essential.

\subsection{Multinucleon transfer to produce superheavy nuclei}
\label{sec:needsmnt}

To complement the fusion studies discussed so far, MNT reactions near the Coulomb barrier are also being considered.  MNT was first proposed as a path to superheavy elements in \cite{Artukh73} and significant work to study this interesting mechanism has been undertaken since then with  U + U reactions at 7.5 MeV/u \cite{Schadel78} as well as U + Cm reactions \cite{Schadel82} at 10 MeV/u.  The authors concluded that the U projectiles may provide a significant advantage for producing superheavy elements because of the very large mass transfer probability and the high neutron-to-proton ratio which makes it possible to reach neutron-rich areas in the "island of stability" inaccessible to transfer or fusion reactions with lighter projectiles.  Zagrebaev and Greiner \cite{zagrebaev2008} highlighted the importance of $^{208}$Pb shell effects ($N=126$) in the $^{136}$Xe+$^{208}$Pb reaction. 
Subsequently, Watanabe et al.'s seminal experiment \cite{watanabe2015} showed significantly larger $N=126$ isotone cross sections from $^{136}$Xe+$^{198}$Pt compared to spallation.  Several reviews have appeared over the years\cite{Kratz15, Loveland19} and, while differing in details of heavy element yield predictions, conclude that MNT is an attractive option study to produce neutron-rich heavy nuclei.  MNT is now a key topic at accelerator facilities.
On the side of theory, multiple approaches have been employed to provide predictions for MNT cross sections.
These include models like the dinuclear system approach \cite{adamian1997b,adamian2020extend}, ImQMD \cite{zhang2020progress}, TDHF \cite{sekizawa2016,sekizawa2019tdhf,godbey2020}, QSMF \cite{lacroix2014stochastic,Sekizawa2020}, as well as multidimensional dynamical models based on the Langevin equations~\cite{Karpov17, Saiko19, Saiko24}.



To find new pathways to unexplored SHN regions requires studying reaction mechanisms for optimal projectile-target and energy combinations to induce MNT towards the desired SHN. As methods like TDHF typically underestimates mass distribution widths, one must extend approaches to include fluctuations (e.g. TDRPA \cite{simenel2012} and QSMF \cite{lacroix2014stochastic}). Given that TDHF alone is a computationally expensive method, using TDRPA/QSMF calculations to explore the reaction landscape poses a significant computational challenge. 
Despite this expense, it is still unclear whether TDHF offers  reasonable estimates for the transfer probabilities populating nuclei far from stability.
To probe this,  systematic experimental data is needed.

\subsection{Uncertainty quantification}
\label{sec:needsuq}

Robust uncertainty quantification (UQ) is critically needed across all theoretical models, especially those guiding experimental design and interpretation. While acknowledged, systematic UQ studies across the SHN production stages (capture, CN formation, evaporation) are largely missing, partly due to numerical complexity and computational cost. The field should reap the benefits of advances in machine learning and emulators to accelerate progress.

In the meantime, Bayesian analyses in DFT have provided uncertainty quantified functionals~\cite{mcdonnell2015}. The next obvious step is to propagate those uncertainties to the time-dependent framework to obtain uncertainty quantified reaction observables. While this has been done for heavy-ion fusion of light systems~\cite{godbey2022uq}, the added expense of quasifission and multinucleon transfer simulations make this a large-scale computational challenge for SHN.
Equally, global optical potentials for nucleons have been extracted from data using Bayesian statistics (e.g. \cite{kduq}) and those uncertainties have been propagated to a variety of direct reactions (e.g. \cite{smith2024}). Similar studies should be performed to quantify optical model uncertainties in observables relevant for SHN production, including barrier distributions obtained from coupled-channel models and transmission coefficients used in Hauser-Feshbach codes. Furthermore, uncertainties on nuclear level densities and strength functions still need to be quantified and propagated to fission and fusion.

As many of the codes used for SHN production are also used more widely for other physics problems, this presents a broad opportunity for collaboration with existing UQ efforts in nuclear structure and reaction theory. Ambiguities in model inputs significantly impact calculated survival probabilities and the field needs to systematically assess the magnitude of these effects. Including UQ from capture, through CN formation, all the way to evaporation, while acknowledging the complex dynamics-structure interplay, is the ultimate goal for reliable theoretical predictions. Providing theoretical predictions with uncertainty estimates would give experimentalists more confidence in pursuing measurements they believe are achievable within defined resources.

While these proposed theory UQ pipelines have the potential to benefit greatly from precise, systematic studies of a large range of observables, even a few measurements in underexplored areas have the potential to make a big impact.
For fusion reactions, for instance, high-resolution fusion excitation functions are not necessary to reveal enough of a trend to be useful in constraining theoretical models.
Indeed, a few well-placed points near the barrier are enough to provide strong evidence toward the height and width of the underlying fusion barrier.
Even structure properties don't necessarily need to be exhaustively studied.
Through the use of microscopic models that intrinsically correlate different observables, information from measurements of masses or excited states can directly inform radii, deformations, and level densities of the same nucleus.
Even beyond single nuclei, forward propagation of the uncertainties through this pipeline will reveal insights into neighboring nuclei.
Even existence measurements, with no indication of detailed nuclear properties or lifetimes, can provide a strong constraint on the mass surface near the proton and neutron driplines.
The incorporation and propagation of these disparate data sources and associated theoretical uncertainties remains a statistical and computational challenge, but the core outcome of extracting the most information possible from complex, expensive experimental campaigns motivates further investment in this direction.


\section{Needs and promising efforts in experiment}
\label{sec:experiments}

This section outlines experimental advances and needs for understanding nuclear reaction mechanisms leading to SHN.
As discussed in Sec.~\ref{sec:needsuq}, the explosion of interest and capability in principled theoretical uncertainty quantification has given any new data on exotic nuclei an immediate role to play in shaping our understanding of nuclear physics.
While some of the efforts in this section will not directly probe the superheavy region, any precise measurements of nuclear properties away from stability are crucial for model calibration and providing better informed input for studies of superheavy formation and structure.
We omit detailed discussion of SHN spectroscopy or current searches, instead focusing on the path to formation.

\subsection{Case Study: The LBNL Superheavy Element Program}
The LBNL 88-inch Cyclotron has a rich SHE research history, now entering a new era due to equipment and electronics upgrades \cite{gates2022studies}. The FIONA spectrometer enables unique gas-phase chemistry experiments by identifying produced species by mass-to-charge ratio \cite{kwarsick2021assessment}, probing relativistic effects on SHE chemistry. Upcoming plans include a Multi-Reflective Time-Of-Flight (MRTOF) for precision mass measurements (probing binding, island of stability, shell effects) and coupling the MRTOF with laser spectroscopy for atomic state energies. The Berkeley Gas-filled Separator (BGS) \cite{gregorich2013simulation} remains central, separating SHE from byproducts. A new silicon detector, SHREC \cite{golubev2024}, with digital electronics is installed at the BGS focal plane. These upgrades support the program's major effort to resume element discovery in the US, starting with the search for element 120, e.g., via the \nuc{50}{Ti}+\nuc{244}{Pu} reaction \cite{gates2024}.

Theoretical predictions can help identify critical data gaps for benchmarking theories. While experimentalists broadly explore isotopes and properties, certain measurements are more crucial for constraining theory. Information on which properties or isotopes are most valuable for honing models would help focus experimental efforts. Theoretical calculations of shell structures, magic numbers, shapes, and deformations are vital. Applied to accessible nuclei, these can guide searches for unique decay modes (like cluster decay), longer-lived nuclei, isotopes with significantly different decay properties, or intriguing isomeric states.

Optimal production of new SHE requires careful beam, target, and energy selection for maximum rates, often meaning detecting only one event over weeks/months. Small energy deviations dramatically reduce rates. Theoretical predictions for new element cross-sections diverge widely \cite{adamian2009feature,liu2013possibility,nasirov2011effects,siwek2012predictions,zagrebaev2008synthesis,zhang2024predictions,zhu2014production}, sometimes by orders of magnitude for slight reaction changes (e.g., $^{48}$Ca to $^{50}$Ti). Predicted optimal energies can differ by tens of MeV, exceeding feasible experimental ranges. Refining predictions requires theories fully discussing differences from previous work and explaining model distinctions. Crucially, the theoretical community can suggest measurements (masses, fission barriers, specific reaction cross-sections) to constrain predictions. Testing reactions producing known elements (like LBNL investigating \nuc{50}{Ti}+\nuc{244}{Pu} for element 116 \cite{gates2024}) before new element campaigns is vital.

\subsection{Capture $\sigma_{cap}$}

There are various ways in which experiment can elucidate the capture phase. One relates to 
the issue of the optimum energy for fusion. Experimental fusion barrier distributions (e.g., quasi-elastic scattering \cite{tanaka2020}) provide crucial data on the effective internuclear potential, and the coupling effects relevant for capture. These measurements constrain theoretical models predicting optimal beam energy for SHN production and provide important data to link structure and dynamic properties in self-consistent theories. Systematic measurement across energy and projectile-target combinations are desired. 

Experiment can also help in testing the validity of the barrier passing model for fusion.
Recent experiments on MNT prior to or accompanying capture \cite{Cook2023, Colucci2024} highlight limitations of simple barrier passing models, showing significant nucleon exchange even below the barrier. This suggests the ``fusing entities" may differ from the initial nuclei. Systematic experimental study of these pre-equilibrium processes is needed to validate and improve theories.

\subsection{Compound nucleus formation $P_{CN}$}

As discussed before, the captured system must form an equilibrated compound nucleus and not re-separate via fission or quasifission. The probability of forming an equilibrated compound nucleus following capture ($P_{CN}$) is the most uncertain term in the final yield cross section ($\sigma_{ER}$). Theoretical predictions vary widely \cite{Loveland_2013}; experimental determinations are upper limits due to overlapping quasifission and fusion-fission mass distributions \cite{Banerjee2019}. Improved experimental limits on $P_{CN}$ will provide important guidance to theory. Specifically, being able to dissect fusion and quasi-fission is important for direct comparison to models. Formation probabilities and quasifission distributions should be obtained for the same reaction conditions to provide a suitable benchmark for theory. The formation stage is strongly coupled to capture; modeling formation needs knowledge of post-capture configurations, dissipation, and transfer. So comprehensive experimental programs that aim to measure a variety of difference observables would be most useful. 

Measurements of quasifission dynamics (mass equilibration, sticking times, TKEs) provide qualitative $P_{CN}$ information. Quasifission ($\sim 10^{-20}$s) shows broader mass distributions and mass-angle correlations \cite{Hinde2021}. Mass-angle distribution analysis is then key to studying the underlying dynamics driving compound nucleus formation. Recent work by Hinde and collaborators~\cite{Hinde:2022wzg} finds experimental mass angle distributions to be at odds with recent theoretical findings, posing an interesting puzzle for theorists to solve. Furthermore, precise studies like that of Tanaka et al.~\cite{tanaka2021} can offer insights into equilibration timescales when paired with large-scale theoretical analysis.

Experiment can also help in understanding whether quasifission is influenced by the same shell effects as fusion-fission. The role of shell effects in quasifission dynamics is fundamental. Do shells influence reseparation like they influence fission from an equilibrated CN? Experimental searches for shell signatures in quasifission fragment distributions (e.g., Hinde et al. \cite{Hinde:2022wzg}) are crucial for validating theoretical predictions and understanding structure-dynamics interplay, providing benchmarks for theoretical nuclear structure properties.

\subsection{Survival against fission $W_{surv}$}
$W_{surv}$ critically depends on the competition between particle emission (neutrons at high $E^*$) and fission from the excited CN. Experimental efforts measure fission probabilities and excitation functions, comparing them to $\Gamma_n / \Gamma_f$ calculations. Precision mass measurements (e.g., GSI SHIPTRAP, RIKEN MRTOF, LBNL MRTOF, FRIB LEBIT) directly probe binding energies ($B_n$), a key $W_{surv}$ input.
Photo-fission studies can also constrain fission barriers and nuclear level densities at lower excitation energies. SHN are an additional argument to perform these measurements.

\subsection{Multinucleon transfer}
SHN production cross-sections for traditional fusion decrease dramatically with Z, from $\sim 0.5$\,pb for Og \cite{Oganessian15} to estimated tens of fb for $Z=119$ and 120. Traditional fusion struggles to produce neutron-rich isotopes. MNT offers a promising alternative, exchanging nucleons in heavy-ion collisions near the barrier, populating products with relatively low excitation energy and high spins \cite{Corradi09}. It has been used for neutron-rich mid-mass nuclei \cite{Corradi09, Gadea03, Bracco21} and studying quasi-elastic to deep-inelastic transitions \cite{Mijatovic22}.

Global MNT efforts for SHE include thermalizing products for decay, laser, and mass spectroscopy (IGISOL-4 \cite{Moore13}, FRS Ion Catcher \cite{Plass13}, KEK KISS \cite{Hirayama15}) and using magnetic spectrometers like PRISMA, which showed MNT favors neutron-rich actinides \cite{Vogt15}.

At Texas A\&M, experiments using \(^{197}\)Au + \(^{232}\)Th and \(^{238}\)U + \(^{232}\)Th at 7.5 MeV/u explored MNT for neutron-rich heavy nuclei. Initial direct detection attempts downstream of a magnet yielded few potential Z=100 events \cite{Natowitz08,Barbui09}. A strategy detecting high-energy alpha chains from products stopped in a catcher exploits increasing alpha energy with emitter charge \cite{Agbemva15}. Passive (polypropylene) and active (YAP scintillator) catcher experiments detected high-energy alpha particles \cite{Wuenschel13, Wuenschel18}, suggesting heavy emitters. Geiger-Nuttall plots implied very heavy elements with long half-lives, but poor YAP resolution/granularity prevented definitive identification.

These results highlight critical detector requirements for future MNT SHE searches: high spatial and energy resolution to associate alphas with single emitters and determine decay properties, and radiation hardness to withstand harsh environments (e.g., near grazing angles). Current technology faces challenges (diamond hardness/cost \cite{Abbott22}, SiC/GaN availability). Alternative concepts include a TPC as a catcher 
or a backward-angle tracking detector coupled with radiation-hard catchers.
Developing advanced detection systems is crucial for realizing MNT's potential to create new superheavy isotopes. Simultaneously, advancements in theoretical descriptions of MNT will be highly beneficial. 

\subsection{Opportunities for radioactive beams and aligning theory with capabilities}
Rare isotope beam facilities like FRIB offer unique opportunities for SHN research. Measuring reactions involving neutron-rich isotope beams can address open questions on the pathways toward more neutron-rich SHN isotopes, inaccessible with stable beams. Theoretical guidance is especially needed to identify promising RIB/target combinations and optimal energies given the much lower rates expected from RIBs. Very significant experimental challenges include RIB intensities, target availability, and detection efficiency for potentially low cross-sections. Intense technological development, guided by theory and experiment, will be required to make this a reality.

The most promising near-term experiments at current leading facilities like FRIB are likely to be studies of heavy-ion reaction dynamics near the barrier for exotic nuclei to improve our understanding of reaction dynamics more generally. Theory constraints obtained from studying other reaction channels can be propagated through to SHN predictions, building a synergy between general purpose facilities and those focused on superheavy synthesis~\cite{frib2024}.

It is important to align theoretical explorations with current and near-future experimental capabilities. Predictions need benchmarking against feasible experiments. Ideally, theories should focus on nuclear properties or reaction mechanisms testable with current capabilities. When theory ventures beyond current capabilities, crucial developments should still be validated with existing data.

While experimentalists lead technology development,  theory often focuses the efforts on specific, high-impact questions that are just within reach. Aligning advancements in experimental capabilities with theoretical predictions will maximize discovery potential. This alignment requires that theorists and experimentalists  collaborate closely, so that both sides understand what is feasible today and what may be accessible in the future.

\section{Other avenues to be considered}
\label{sec:other}

Given the difficulties faced with current experimental approaches to SHN synthesis, we wanted to include a section that discusses other, potentially more speculative, avenues to stimulate new ideas for SHN production. In this section, we discuss the opportunity offered by cluster transfer into doorway states in the continuum (Sec.~\ref{sec:otherdoorway}),  the possibility of taking advantage of a neutron-rich environment to generate a succession of neutron captures (Sec.~\ref{sec:otherncapture}), and the potential for SHN production in astrophysical sites (Sec.~\ref{sec:otherastro}).

\subsection{Doorway states}
\label{sec:otherdoorway}

The synthesis of heavy and superheavy isotopes through direct fusion of binary systems faces significant challenges, primarily due to the competing factors of bombardment energy and survival probability after the formation of compound nuclei, as discussed throughout this paper. When the incoming energy is lower than the Coulomb barrier, preventing the target and projectile from getting close enough to form a compound nucleus, the fusion cross-section is extremely small. However, at these low energies, the fission barrier is higher, resulting in a higher survival probability, which can lead to the formation of heavy and superheavy isotopes. Conversely, when the incoming energy is high and exceeds the Coulomb barrier, the fusion cross-section increases. Unfortunately, the excitation energy of the compound nuclei also surpasses the fission barrier, leading to a lower survival probability. This, in turn, decreases the overall cross-section for synthesizing heavy and superheavy isotopes.

The Trojan Horse method (THM) may offer a potential solution to overcome these challenges. This indirect approach involves forming the same compound nuclei through a different reaction pathway. Consider a projectile with a two-body structure, represented as $ a = b + x $, where $ a $ is the projectile, and $ b $ and $ x $ are its components. These components are relatively loosely bound within the projectile, facilitating the breakup of $ a $ into $ b $ and $ x $.

In a reaction where the projectile $ a $ interacts with a target nucleus $ A $, the following channel can occur:

\begin{equation}
a(=b+x) + A \to b + B^*,
\end{equation}
where $ B^* $ is the compound nucleus that can be formed by the direct reaction of $ x + A $. 

In a semi-classical framework, the reaction pathway can be described as follows: first, the projectile $ a $ overcomes the Coulomb barrier between $ a $ and $ A $ if the energy of $ a $ is sufficiently high. Subsequently, the projectile $ a $ breaks into $ b $ and $ x $. The component $ x $ then interacts with the target nucleus $ A $ to form the compound nucleus $ B^* $, while the fragment $ b $ escapes the interaction region.

In this process, since $ x $ is already inside the Coulomb barrier—brought in by the Trojan Horse projectile $ a $—it does not experience the full Coulomb barrier that $x$ would face in a direct reaction with $A$. This can lead to an effectively higher fusion cross-section at energies that would be below the Coulomb barrier in the direct $x+A$ reaction. Additionally, with the high fission barrier at these lower effective energies, the formation probability of superheavy isotopes is potentially increased.

The theoretical model to study the Trojan Horse process can be traced back to the work by Ichimura, Austern, and Vincent (IAV) in the 1980s, where they discussed the inclusive breakup process within a three-body model \cite{Ichimura1985}. They identified two main mechanisms for inclusive breakup: elastic breakup and nonelastic breakup (NEB).

In elastic breakup, after the collision, both fragments interact elastically with the target nucleus $ A $. In nonelastic breakup, at least one fragment interacts nonelastically with the target. This includes processes such as particle exchange between $ x $ and $ A $, inelastic scattering leading to excited states of $ x $ or $ A $, and the fusion process of $ x $ with $ A $ to form a compound nucleus.

In the IAV model, the NEB cross-section can be expressed as:
\begin{equation}
\left. \frac{d^2\sigma}{dE_b d\Omega_b} \right|_\mathrm{NEB} = -\frac{2}{\hbar v_{a}} \rho_b(E_b) \langle \varphi_x (\mathbf{k}_b) | \mathrm{Im}[U_{xA}] | \varphi_x (\mathbf{k}_b) \rangle,
\end{equation}
where $ \rho_b(E_b) $ is the density of states of particle $ b $, $ v_a $ is the velocity of the incoming particle $a$, and $ \varphi_x(\mathbf{k}_b, \mathbf{r}_{xA}) $ is a relative wave function describing the motion between $ x $ and $ A $ when particle $ b $ is scattered with momentum $ \mathbf{k}_b $.

The wave function $ \varphi_x(\mathbf{k}_b, \mathbf{r}_{x}) $ is obtained from the equation:
\begin{equation}
\varphi_x(\mathbf{k}_b, \mathbf{r}_{x}) = \int G_x (\mathbf{r}_{x}, \mathbf{r}'_{x}) \langle \mathbf{r}'_{x} \chi_b^{(-)}(\mathbf{k}_b) | \mathcal{V} | \Psi^{3b(+)} \rangle d\mathbf{r}'_{x},
\end{equation}
where $ G_x $ is the Green's function with optical potential $ U_{xA} $, $ \chi_b^{(-)*}(\mathbf{k}_b, \mathbf{r}_{b}) $ is the distorted wave describing the relative motion between $ b $ and the $ B^* $ compound system (obtained with some optical potential $ U_{bB} $), $ \mathcal{V} \equiv V_{bx} + U_{bA} - U_{bB} $ is the post-form transition operator, and $ \Psi^{3b(+)} $ is the three-body scattering wave function.

When heavy ions collide near the Coulomb barrier, multiple reaction channels become available, with Q-values for each channel primarily dictating the preferred pathways. For weakly bound projectile systems, breakup followed by partial fusion has been shown to constitute an important component of the reaction mechanism. In reactions where favorable Q-value matching exists, sequential capture of the two clusters can become more probable than total fusion, owing to the lower Coulomb barrier encountered by the fragments. This mechanism may provide a novel doorway for superheavy nuclei (SHN) production. When the heavy projectile exhibits significant clusterization, partial fusion can emerge as a dominant pathway, with its importance ultimately determined by the Q-values associated with the various competing mechanisms.

Studies utilizing halo nuclei~\cite{Jin25} have demonstrated that the Trojan Horse effect significantly influences the nonelastic breakup cross-section, leading to a substantial increase in the cross-section below the Coulomb barrier compared to direct reactions. This enhancement arises from the weak binding characteristic of halo nuclei, which significantly enhances sub-barrier fusion cross sections. This phenomenon opens promising pathways for synthesizing new elements, including the potential use of giant halo nuclei to explore uncharted territories in the periodic table.

\subsection{Using successive n-capture}
\label{sec:otherncapture}

One nonstandard method explored in the mid-twentieth century is the production of heavier elements through successive neutron captures. The resulting short-lived nuclei in turn transmute to nuclei with larger $Z$ via nuclear $\beta$-decay.
This process is similar to the well-known astrophysical rapid neutron capture process or $r$-process of nucleosynthesis.

In fact, it was the early study of the debris of nuclear explosions in the 1950s that led to the idea of the $r$-process \cite{Cameron1957, Burbidge1957}. Later, the US Heavy Element Program, part of Project Plowshare, aimed to investigate the production of heavy elements through successive neutron captures in peaceful nuclear explosions. At the time, both Fermium and Einsteinium were new elements discovered in the debris of nuclear tests. Subsequent exploration of over 30 tests did not yield evidence of any heavier elements being produced beyond $^{257}$Fm \cite{Becker2016}.

The limitation of this approach is that prevalent neutrons in dense environments also trigger fission in heavy nuclei. Neutron-induced fission converts heavy nuclei into lighter products, thereby terminating the required march along the isotopic chain of the heavy target. Any nuclear chain that encounters low fission barriers will stop the successive neutron-capture process (or any process for that matter), and there will not be sufficient quantities of neutron-rich nuclei that would eventually $\beta$ decay to heavier elements. Even if sufficient neutron capture could occur, it is likely to be limited at extreme neutron excess by $\beta$-delayed fission \cite{Mumpower2022} or by spontaneous fission \cite{Guiliani2018}.

Successive neutron captures may be produced via other means. In the past decade, it has been proposed that upcoming multi-Petawatt, few-femtosecond laser systems could be of interest in producing neutron beams \cite{Roth2013}. The mechanism for neutron production involves the acceleration of electrons or ions from varying targets. Bremsstrahlung $\gamma$-rays or nuclear fusion of protons with light nuclei trigger the release of neutrons \cite{Chen2019}. These high-intensity laser systems could be used in the future to produce neutron-rich nuclei relevant for the study of nucleosynthesis processes \cite{Hill2021}. Such studies would not be limited to particular nuclei, and thus could be applied to the study of superheavy nuclei in a less destructive manner than the US Heavy Element program. Future investigations in this regard could shed light on fission barriers of extremely neutron-rich short-lived species, and may give more insights into superheavy nuclei if production could be achieved. It has been pointed out that the areal density of the neutron source is currently too low with present techniques and that this crucial quantity may need to be increased several orders of magnitude for these methods to be feasible \cite{Horny2024}.

\subsection{Possible synthesis in extreme stellar environments}
\label{sec:otherastro}

So far, there is no astrophysical observation of  SHN production in r-process nucleosynthesis \cite{Holmbeck2023b}. One difficulty is the presence of nuclei with low fission barriers which prevents the r-process path from reaching the beta stability line in the superheavy region, as discussed in the previous Section.

Recently, it has been pointed out that superheavy nuclei may emerge as an equilibrium composition in the outer crust of neutron stars with extremely large magnetic fields ($B\gtrsim 10^{18}$\,G) \cite{yoshimura2025,basilico2024}.
Under these extremely high magnetic fields, superheavy nuclei with neutron numbers $N\simeq260$--$290$ may emerge as an equilibrium composition in the deepest layers of the outer crust. These neutron numbers are affected by the shell closure after $N=184$.

The predicted magnetic field strength of $B\gtrsim10^{18}$\,G is stronger than the currently inferred surface magnetic fields of magnetars (around $B=10^{14\text{--}15}$\,G). However, it has been argued that exotic scenarios such as toroidal magnetic fields, where $B\approx10^{18}$\,G can be reached locally, or other dynamical origins like neutron star mergers may trigger superstrong magnetic fields \cite{naso2008,frieben2012}. Moreover, when magnetic fields exceed $B\simeq10^{17}$\,G, they affect not only electron energies but also nuclear single-particle energy levels and, thus, nuclear structure \cite{arteaga2011,Basilico2015,Stein2016,Jiang2024}.
Considering the effect of superstrong magnetic fields in microscopic calculations of superheavy nuclei within a variety of theoretical formulations is important to explore possible syntheses of those nuclei under extreme stellar conditions.

\section{Building Community}
\label{sec:community}

 SHN experiments are extremely challenging and rely heavily on theoretical predictions. For theory to fully play its role in SHN research, a strong and sustainable community is needed. While there are strong experimental efforts in various parts of the world, the community of theorists working specifically on SHN production problems is small.

Modeling the production of SHN is a monumental task that requires the concerted efforts of a dedicated theory community with diverse expertise. Some nuclear structure theorists are interested in heavy nuclei and, as a consequence of the tools developed in their nuclear structure research, are able to make valuable contributions to the field of SHN.

However, developing new techniques and investing the person-power needed for novel implementations in SHN production requires theory programs dedicated to the topic. Until such a theory community is built, theory cannot provide the best guidance to experiment. The situation will worsen if the next generation of theorists are not brought into the field. A critical mass of senior theorists with research programs dedicated to SHN production is essential to train the early career individuals who will ensure future success.

Recognizing the situation of the SHN theory community, one of the goals of the FRIB-TA topical program was precisely to serve as a catalyzer toward building a strong theory community with the diverse expertise needed. The complexity and scale of the work to be done can benefit from international collaboration, pooling together resources, expertise, and unique perspectives from across the globe. By working together, this community can push the boundaries of our knowledge, uncovering new elements and isotopes, and refining our understanding of nuclear physics. Building and sustaining this community requires deliberate effort, coordinating research,  fostering collaboration, facilitating data sharing, while developing the next generation of SHN scientists.


\section{Conclusions}

The synthesis and study of superheavy nuclei represent a frontier in nuclear physics, chemistry, and astrophysics, probing the limits of nuclear existence and providing stringent tests for theoretical models. Despite significant progress in discovering elements up to $Z=118$, accessing more neutron-rich isotopes in the predicted SHN island (or peninsula) of stability and discovering elements beyond $Z=118$ remains extremely challenging due to rapidly decreasing production cross-sections and short half-lives.

This report summarizes the discussions from the 2024 FRIB-TA topical program, identifying key challenges and opportunities. On the theoretical front, significant advancements are needed across all stages of SHN production, from the initial capture to the eventual de-excitation of a compound nucleus.
Experimentally, progress requires not only pushing the boundaries of beam intensity and detection efficiency but also carefully planning measurements specifically designed to constrain theoretical models.
This tight feedback loop also necessitates advances in computing and data science methodology, enabling theory predictions to cover a broad range of reactions systematically including a statistically relevant uncertainty.

Ultimately, realizing the potential of SHN research requires a strong, collaborative interplay between theory and experiment. Theory must provide predictive power rooted in fundamental physics while being benchmarked against experimental data. Experiment, in turn, must generate targeted data to test and refine theoretical predictions. This collaborative effort, spanning diverse expertise and international boundaries, is essential. A critical challenge is building and sustaining a dedicated theoretical community for SHN, including training the next generation of scientists to address the complex, multi-faceted problems in this exciting and demanding field. 

\section*{Acknowledgements}

The authors would like to thank the FRIB Theory Alliance for hosting the Topical Program. Authors' individual acknowledgments follow:

K.G. and W.N. acknowledge support by the U.S. Department of Energy under Award Numbers DE-SC0013365, DE-NA0004074 (NNSA, the Stewardship Science Academic
Alliances program), and DE-SC0023175 (Office of Science,
NUCLEI SciDAC-5 collaboration).
The work of F.M.N. was in part supported by the U.S. Department of Energy grant DE-SC0021422 and National Science Foundation CSSI program under award No. OAC-2004601 (BAND Collaboration).
K.J.C. acknowledges support by the Australian Research Council Grant DE230100197.
J.M.G. and J.L.P. were supported by the U.S. Department of Energy, Office of Science, Office of Nuclear Physics under contract number DE-AC02-05CH11231. 
K.Hagel was supported by the U.S. Department of Energy Grant No. DE-FG02-93ER40773.
K.Hagino acknowledges support by JSPS KAKENHI Grant Number JP23K03414.
M. Kowal was partially co-financed by the COPIGAL Project.
J.Lei acknowledge support by the National Natural Science Foundation of China under Grants No. 12475132
J.Lubian and J.R. were partially supported by CNPq, FAPERJ, and Instituto Nacional de Ciência e Tecnologia- Física Nuclear e Aplicações (Project No. 464898/2014-5).
A.M. was supported by the Polish National Science Center Grants under Contract No. UMO- 2021/43/B/ST2/01191.
M.R.M. was supported by the U.S. Department of Energy through the Los Alamos National Laboratory. 
Los Alamos National Laboratory is operated by Triad National Security, LLC, for the National Nuclear Security Administration of U.S.\ Department of Energy (Contract No.\ 89233218CNA000001).
K.S. acknowledges support by  JSPS Grant-in-Aid for Scientific Research, Grants No. 23K03410 and No. 23K25864.
A.S.U. acknowledges support by the U.S. Department of Energy under award number DE-SC0013847 (Vanderbilt University).

\section*{References}
\bibliography{she}
\end{document}